\documentclass[preprint,12pt]{elsarticle}

\usepackage{graphicx}
\usepackage{epstopdf}
\usepackage{multirow}

\usepackage{amssymb}

\journal{Journal of Physics G}

\begin{document}

\begin{frontmatter}



\title{Feasibility of studying the K$^*_0(700)$ resonance using $\pi^{\rm \pm}$K$^0_{\rm S}$ femtoscopy}


\author{T. J. Humanic}

\address{Department of Physics, Ohio State University, Columbus, OH, USA}

\begin{abstract}
The feasibility of using $\pi^{\rm \pm}$K$^0_{\rm S}$ femtoscopy to experimentally study the K$^*_0(700)$ resonance is considered. The K$^*_0(700)$ resonance is
challenging to study due to its broad width and proximity in mass to the K$^*(892)$, both
of which predominantly decay via the $\pi$K channel. One of the main interests in 
the K$^*_0(700)$ is that it is considered a candidate for a tetraquark state. 
It is proposed to use two-particle femtoscopic methods with $\pi^{\rm \pm}$K$^0_{\rm S}$
pairs produced in proton-proton and heavy ion collisions assuming a strong final-state 
interaction between them due to elastic scattering and/or via the K$^*_0(700)$ resonance. Calculations of $\pi^{\rm \pm}$K$^0_{\rm S}$ correlation functions are made to
estimate the strength of these final-state interactions and to see if this signal can be 
adequately separated from the K$^*(892)$ background. A possible signature of
diquark versus tetraquark behavior of the K$^*_0(700)$ final-state interaction is
also discussed.
\end{abstract}



\end{frontmatter}


\section{Introduction}
The strange K$^*_0(700)$ meson is listed in the 2018 Review of Particle Physics (RPP) with the
qualifier ``Needs confirmation''~\cite{Tanabashi:2018oca}. Selected primary information known about it
from the RPP is: isospin, $I=1/2$, average mass,  $m_{\rm (700)}= 824\pm30$ MeV/$c^2$, and
average width, $\Gamma_{\rm (700)} = 478\pm50$ MeV/$c^2$. This information is mostly obtained from experiments 
that study heavy particle decays into the kaon-pion channel, which appears to be its primary decay
channel as is also the case for the well-studied 
K$^*(892)$ which is nearby in mass~\cite{Ablikim:2010ab,Epifanov:2007rf}. The broad width
of the K$^*_0(700)$ along with its proximity to the much narrower K$^*(892)$ make it challenging
to study since it appears as a ``shoulder'' of the K$^*(892)$ in the kaon-pion invariant mass 
distribution.
The main interest in the K$^*_0(700)$ is that its isospin, mass and decay channel fit nicely in
the low-mass tetraquark nonet that has been predicted~\cite{Jaffe:1976ig,Alford:2000mm}.
However, no direct and convincing evidence of its tetraquark nature currently exists.

The ALICE collaboration at the Large Hadron Collider has endeavored to experimentally
probe the quark nature of another of the mesons in the tetraquark nonet, the $a_0(980)$,
using K$^0_{\rm S}$K$^{\rm \pm}$ femtoscopy in pp collisions at $\sqrt{s}=7$ TeV~\cite{Acharya:2018kpo} and Pb-Pb collisions at $\sqrt{s_{\rm NN}}=2.76$ TeV~\cite{Acharya:2017jks}.
The idea was to study the $a_0(980)$ as the final-state interaction (FSI) between the 
kaons in the K$^0_{\rm S}$K$^{\rm \pm}$ pair, the strength of the FSI being related to
the quark nature of the $a_0(980)$. The authors conclude that
their results suggest that the $a_0(980)$ is a tetraquark state.

The purpose of the present paper is to determine the feasibility of using femtoscopic methods
similar to the ones used by the ALICE collaboration mentioned above in pp and heavy-ion
collisions to experimentally study the K$^*_0(700)$ via the $\pi^{\rm \pm}$K$^0_{\rm S}$
channel. The $\pi^{\rm \pm}$K$^0_{\rm S}$ channel is used so that no final-state Coulomb
interaction is present between the particles.
The $\pi^{\rm \pm}$K$^0_{\rm S}$
correlation function will be calculated assuming an elastic FSI and a FSI through the 
K$^*_0(700)$, both in the presence of the K$^*(892)$, to determine the strength of the
FSI and the effect on the FSI signal of the K$^*(892)$ background. 
A possible signature of
diquark versus tetraquark behavior of the K$^*_0(700)$ final-state interaction will
also be mentioned.

\section{Calculational methods}
The $\pi^{\rm \pm}$K$^0_{\rm S}$ correlation function calculated in the present work, $C(k^*)$,
is the sum of the FSI contribution, $C_{\rm FSI}(k^*)$, and K$^*(892)$ contribution, $C'_{\rm (892)}(k^*)$, otherwise assuming a flat baseline,
\begin{equation}
C(k^*)=C_{\rm FSI}(k^*)+C'_{\rm (892)}(k^*)
\label{cf}
\end{equation}
where $k^*$ is the momentum of one of the particles in the pair reference frame.
The calculation of these contributions is now separately described below.

\subsection{Final-state interaction contribution to the correlation function}
As was done by the ALICE collaboration in their $a_0(980)$ 
studies~\cite{Acharya:2018kpo,Acharya:2017jks},
the femtoscopic FSI contribution, $C_{\rm FSI}(k^*)$, is calculated with the correlation function from
R. Lednicky and is based on the model of R. Lednicky and V.L. Lyuboshitz~\cite{Lednicky:1981su,Lednicky:2005af} (see also Ref.~\cite{Abelev:2006gu}).
\begin{equation}
C_{\rm FSI}(k^*)=1+\lambda\alpha\left[\frac{1}{2}\left|\frac{f(k^*)}{R}\right|^2+\frac{2\mathcal{R}f(k^*)}{\sqrt{\pi}R}F_1(2k^* R)-\frac{\mathcal{I}f(k^*)}{R}F_2(2k^* R)\right],
\label{cf_fsi}
\end{equation}
where
\begin{equation}
F_1(z)\equiv\frac{\sqrt{\pi} e^{-z^2} {\rm erfi}(z)}{2 z};\qquad F_2(z)\equiv\frac{1-e^{-z^2}}{z}
\label{eq:fit3}
\end{equation}
where $f(k^*)$ is the scattering amplitude, $\alpha$ is the fraction of $\pi^\pm$K$^{0}_{\rm S}$ pairs that come 
from the $\pi^-$K$^0$ or  $\pi^+$${\rm \overline{K}^0}$ system, set to 0.5 assuming symmetry in K$^0$ and 
${\rm \overline{K}^0}$ production \cite{Abelev:2006gu}, $R$ is the radius parameter assuming a
spherical Gaussian source distribution, and $\lambda$ is the correlation strength. The correlation strength is
unity in the ideal case of pure FSI, which is assumed in the present work.

Correlation functions are calculated using two different scattering amplitudes: resonant 
scattering through the K$^*_0(700)$, and elastic scattering.

\subsubsection{Resonant scattering through the K$^*_0(700)$}
The equation used for the scattering amplitude for resonant scattering through the
K$^*_0(700)$ was adapted from Eq.~11 in Ref.~\cite{Abelev:2006gu},
\begin{equation}
f(k^*) = \frac{\gamma}{m_{\rm (700)}^2-s-i\gamma k^*}
\label{sa_kstar0}
\end{equation}
where, $\gamma=m_{\rm (700)}\Gamma_{\rm (700)}/k_{\rm (700)}$, $k_{\rm (700)}$
is the decay momentum of a daughter in the pair frame in the K$^*_0(700)$ decay, and
$s=(\sqrt{k^{*2}+m_\pi^2}+\sqrt{k^{*2}+m_{K^0}^2})^2$. The RPP values stated earlier were
used for $m_{\rm (700)}$ and $\Gamma_{\rm (700)}$, and $k_{\rm (700)}=235.3$ MeV/$c$.

\subsubsection{Elastic scattering}
For the FSI elastic scattering into isospin state $I$, the scattering amplitude equation
for s-wave and small $k^*$ was used~\cite{Pelaez:2016tgi},
\begin{equation}
f_I(k^*)=\frac{-a_I}{1+ia_Ik^*}
\label{sa_elastic}
\end{equation}
where $a_I$ is the scattering length for isospin state $I$. Since there are two possible
isospin states for pion-kaon elastic scattering, $I=1/2$ and $I=3/2$, 
the scattering
amplitude used in Eq.~\ref{cf_fsi} was taken as the average of the two isospin 
states~\cite{Abelev:2006gu},
\begin{equation}
f(k^*)=\frac{f_{1/2}(k^*)+f_{3/2}(k^*)}{2}
\label{sa_average}
\end{equation}
where the pion-kaon scattering lengths were taken from 
Ref.~\cite{Pelaez:2016tgi}, $m_{\pi}a_{1/2}=0.22$
and $m_{\pi}a_{3/2}=-0.054$.

\subsection{K$^*(892)$ contribution to the correlation function}
The contribution to the $\pi^{\rm \pm}$K$^0_{\rm S}$ correlation function from the
decay of the K$^*(892)$ was calculated assuming a non-relativistic Breit-Wigner
function for the pion-kaon invariant mass distribution, $dN/dm$~\cite{Abelev:2012hy},
\begin{equation}
\frac{dN}{dm}\propto\frac{1}{2\pi}\frac{\Gamma_{\rm (892)}}{(m-m_{\rm (892)})^2+\Gamma_{\rm (892)}^2/4}
\label{bw}
\end{equation}
where $m_{\rm (892)}=891.76$ MeV/$c^2$ and $\Gamma_{\rm (892)}=50.3$ MeV/$c^2$
are the mass and width of the K$^*(892)$ from the RPP. The contribution to the correlation
function from this in terms of $k^*$ can be written as
\begin{equation}
C'_{\rm (892)}(k^*)=B\frac{dN}{dm}\frac{dm}{dk^*}=B\frac{dN}{dm}k^*\left(\frac{1}{\sqrt{k^{*2}+m_\pi^2}}+\frac{1}{\sqrt{k^{*2}+m_{K^0}^2}}\right)
\label{bw_cf}
\end{equation}
where $B$ is a normalization factor. An approximate value for $B=7.37\times10^{-3}$ GeV was obtained from the
ALICE experiment that measured K$^*(892)^0 \rightarrow \pi^\pm K^\mp$ production 
in 7 TeV pp collisions by estimating the ratio of the unlike-charged particles
peak in the vicinity of the K$^*(892)^0$ mass to the mixed-event background in the 
raw invariant mass distribution,
Fig.~3 in Ref.~\cite{Abelev:2012hy}. The uncertainty in the scale of $C'_{\rm (892)}(k^*)$
resulting from this procedure is estimated to be $\pm15\%$, which is judged to be sufficient for
the present feasibility study.

\section{Results}
\begin{figure}
\begin{center}
\includegraphics[width=120mm]{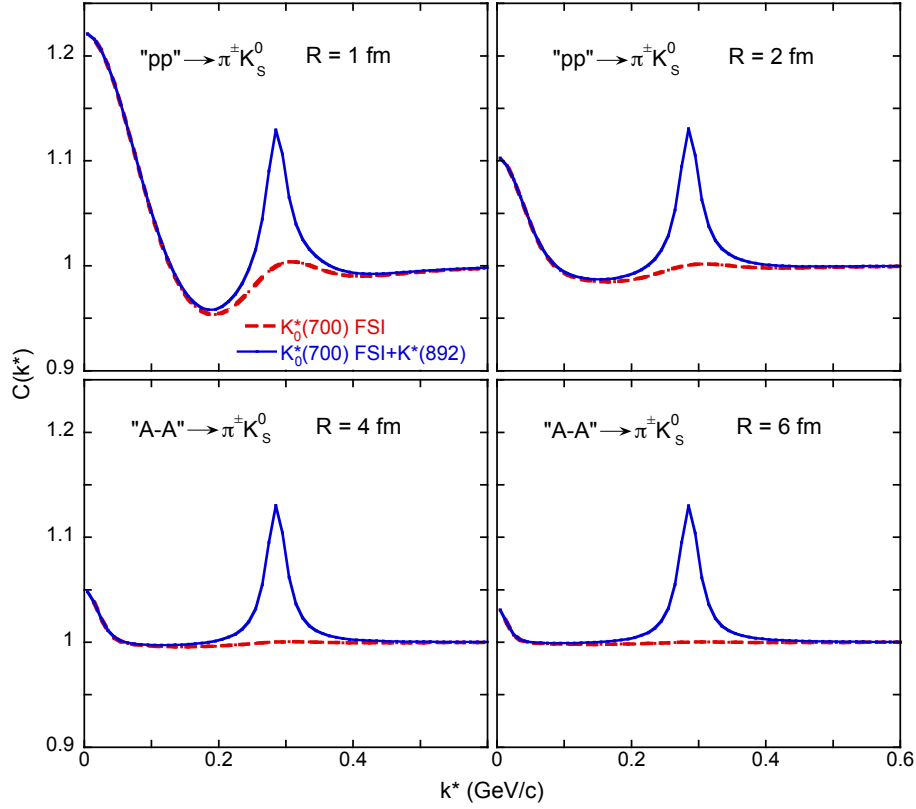} \caption{$C(k^*)$ with the resonant K$^*_0(700)$
FSI for four $R$ values. The solid (blue) line is the full calculation and the dashed (red) line shows
the K$^*_0(700)$ FSI component only (color online).}
\label{fig1}
\end{center}
\end{figure}

\begin{figure}
\begin{center}
\includegraphics[width=120mm]{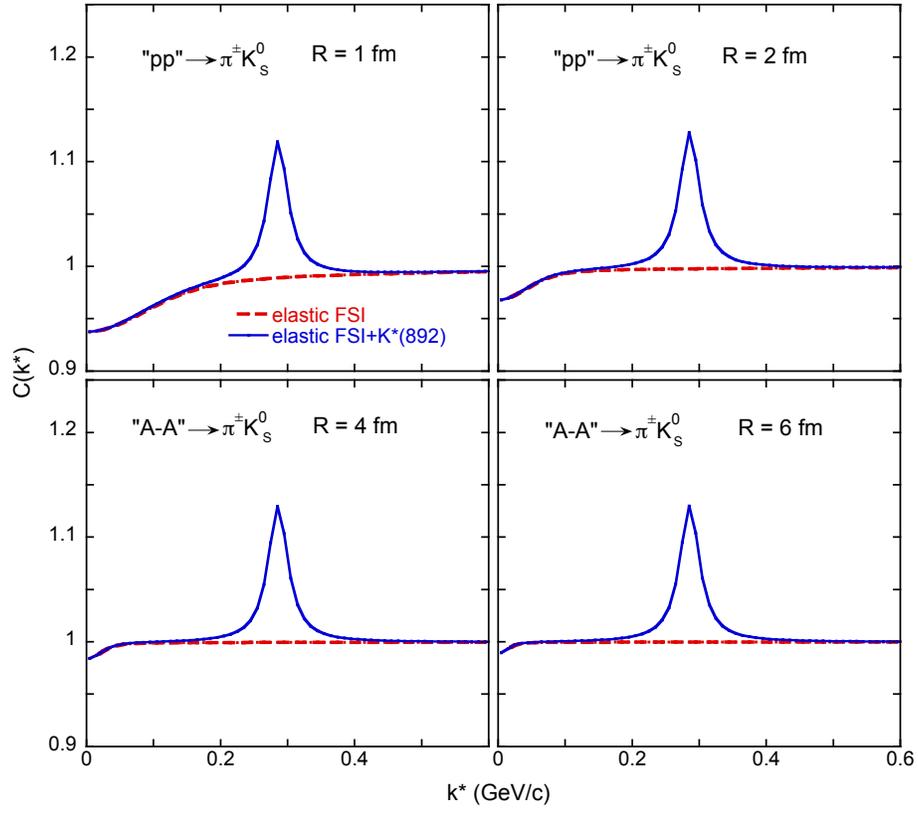} \caption{$C(k^*)$ with the elastic scattering
FSI for four $R$ values. The solid (blue) line is the full calculation and the dashed (red) line shows
the elastic FSI component only (color online).}
\label{fig2}
\end{center}
\end{figure}

Figures~\ref{fig1} and \ref{fig2} show the results for $C(k^*)$ from calculations with Eq.~\ref{cf}
for the K$^*_0(700)$ resonant FSI and the elastic scattering FSI, respectively. Calculations are shown
in each figure for four values of the radius parameter, $R=$1, 2, 4 and 6 fm. The $R=$ 1 and 2 fm
calculations represent the typical source sizes measured by femtoscopy in pp collisions and 
the $R=$ 4 and 6 fm calculations represent those typically measured in heavy-ion collisions.
The solid (blue) lines represent the full calculation whereas the dashed (red) lines represent
the contribution from $C_{\rm FSI}(k^*)$ only, i.e. Eq.~\ref{cf_fsi}. In each figure the 
peak of the K$^*(892)$ contribution is located at $\sim 0.3$ GeV/$c$.
Although the scale of
$C'_{\rm (892)}(k^*)$ was set using the results of a 7 TeV pp collisions experiment as described
above, it is not expected to change significantly with bombarding energy or 
colliding species~\cite{Adam:2017zbf}.

In Fig.~\ref{fig1} the resonant K$^*_0(700)$ FSI is seen to give an enhancement near
$k^*\sim 0$ that is larger or comparable to the K$^*(892)$ contribution for $R=$ 1 and 2 fm
and then decreasing in size as $R$ increases. It is also seen that the
FSI signal overlaps into the region of the K$^*(892)$ for the smaller values of $R$ but is well-separated from the K$^*(892)$ for $R=$ 4 and 6 fm.

The elastic scattering FSI used in the calculation of Fig.~\ref{fig2} gives a qualitatively different
signal than seen in Fig.~\ref{fig1} in that it produces a depletion near $k^*\sim 0$ of smaller
magnitude than the enhancement produced in the resonant K$^*_0(700)$ FSI case. As also seen
for the resonant K$^*_0(700)$ FSI case, the magnitude of the elastic scattering signal reduces
with increasing $R$ and becomes better separated from the K$^*(892)$ peak.

With these results, one can now try to evaluate the feasibility of carrying out
$\pi^{\rm \pm}$K$^0_{\rm S}$ femtoscopic experiments to study the K$^*_0(700)$ in pp
and/or heavy-ion collisions. It is assumed that an experimental $\pi^{\rm \pm}$K$^0_{\rm S}$
correlation function can be constructed that is corrected for non-flat baseline effects,
such as with a Monte Carlo calculation. Looking at Fig.~\ref{fig1} and \ref{fig2}, it should
be possible to distinguish between the two FSI cases since they have qualitatively different
shapes. Also, the FSI signals are reasonably well-separated from the K$^*(892)$ peak. If it can be assumed that these are the only two FSI possibilities, a correlation function
similar to Eq.~\ref{cf} that combines both FSI cases could be fit to the experimental one, such as
\begin{equation}
C(k^*)=1+\delta[C^{\rm res}_{\rm FSI}(k^*)-1]+(1-\delta)[C^{\rm el}_{\rm FSI}(k^*)-1]+C'_{\rm (892)}(k^*)
\label{cf_both}
\end{equation}
where $C^{\rm res}_{\rm FSI}(k^*)$ and $C^{\rm el}_{\rm FSI}(k^*)$ are the resonant K$^*_0(700)$
and elastic FSI versions of Eq.~\ref{cf_fsi}, and $\delta$ is a fit parameter that gives the fraction
of each FSI case, i.e. for $\delta=1$ only the resonant K$^*_0(700)$ FSI is present. The same arguments as used by the ALICE collaboration in considering
whether or not the $a_0(980)$ is a tetraquark state can be applied to the K$^*_0(700)$. 
If $\delta=1$ for heavy-ion collisions, this would be consistent with a tetraquark K$^*_0(700)$
since it implies a direct transfer of the quarks in the $\pi^{\rm \pm}$K$^0_{\rm S}$ system
to the K$^*_0(700)$, which is facilitated by the large size of the heavy-ion collisions interaction
region since it is less likely for a $d\overline d$ annihilation. For pp collisions, the pion and
neutral kaon are produced in close proximity to each other enhancing the probability
for a $d\overline d$ annihilation that would compete with tetraquark formation but
enhance a diquark K$^*_0(700)$. Thus for the pp collision case $\delta=1$ would
be consistent with a diquark K$^*_0(700)$ whereas, $\delta<1$ would be consistent with
a tetraquark K$^*_0(700)$ (See
Refs.~\cite{Acharya:2018kpo,Acharya:2017jks} for a more detailed presentation of
these arguments).

On a technical note, it should be recalled that the results shown in Fig.~\ref{fig1}
and \ref{fig2} assume that the FSI correlation strength, $\lambda$ in Eq.~\ref{cf_fsi}, is unity.
Normally in femtoscopic experiments one measures $\lambda<1$ due to the presence of
long-lived resonances and non-Gaussian sources~\cite{Acharya:2017jks}, 
resulting in a smaller
signal size than shown in the figures. Nevertheless, 
femtoscopic experiments routinely measure signals of this size or less, 
e.g. see Refs.~\cite{Acharya:2018kpo,Acharya:2017jks}, so this should not be a problem.
 
\section{Summary}
The feasibility of using $\pi^{\rm \pm}$K$^0_{\rm S}$ femtoscopy to experimentally study the K$^*_0(700)$ resonance was considered. The $\pi^{\rm \pm}$K$^0_{\rm S}$
correlation function was calculated assuming an elastic FSI and a FSI through the 
K$^*_0(700)$, both in the presence of the K$^*(892)$, to determine the strength of the
FSI and the effect on the FSI signal of the presence of the K$^*(892)$ background. 
It was found that the resonant K$^*_0(700)$ FSI is large and qualitatively different in
shape compared with the elastic scattering FSI, making it possible to distinguish between
the two. It was also found that the resonant K$^*_0(700)$ FSI signal is sufficiently well
separated from the K$^*(892)$ peak in the correlation function. It is thus judged to be
feasible to experimentally carry out such a study.
A possible signature of
diquark versus tetraquark behavior of the K$^*_0(700)$ final-state interaction
that could be applied in such an experiment was
also mentioned.





The author wishes to acknowledge financial support from the U.S.
National Science Foundation under grant PHY-1614835.

\bibliographystyle{utphys}   
\bibliography{K0pi.bib}

\end{document}